\documentclass[preprint,floatfix] {revtex4} 
\newcommand{\rvec}{\mathrm {\mathbf {r}}} 
 
\newcommand{\Rvec}{\mathrm {\mathbf {R}}}

\newcommand{\kvec}{\mathrm {\mathbf {k}}}

\newenvironment{rcases}
  {\left.\begin{aligned}}
  {\end{aligned}\right\rbrace}
  
\usepackage{graphicx}
\usepackage{subfigure}
\usepackage{xcolor}
\usepackage{enumerate}
\usepackage{braket}
\usepackage{amsmath, bm}
\usepackage{stackengine}

\usepackage{color, soul}
\definecolor{darkblue}{rgb}{0,0,0.5}
\setulcolor{darkblue}

\begin{document}

\title{Excitation energies through Becke's exciton model within a Cartesian-grid KS DFT}

\author{Abhisek Ghosal, Tarun Gupta, Kishalay Mahato}
\author{Amlan K.~Roy}
\altaffiliation{Email: akroy@iiserkol.ac.in, akroy6k@gmail.com.}                                           
\affiliation{Department of Chemical Sciences\\
Indian Institute of Science Education and Research (IISER) Kolkata \\  
Nadia, Mohanpur-741246, WB, India}

\begin{abstract}
Photon-induced electronic excitations are ubiquitously observed in organic chromophore. In this context, we present a simple, alternative 
time-independent DFT procedure, for computation of single-particle excitation energies, in particular, the lower bound excited singlet 
states, which are of primary interest in photochemistry. This takes inspiration from recently developed Becke's exciton model, where a key 
step constitutes the accurate evaluation of correlated singlet-triplet splitting energy. It introduces a non-empirical model, both from 
``adiabatic connection theorem" and ``virial theorem" to analyze the role of 2e$^-$ integral in such calculations. The latter quantity is 
efficiently mapped onto a real grid and computed accurately using a purely numerical strategy. Illustrative calculations are performed on 
10 $\pi$-electron organic chromophores within a Cartesian-grid implementation of pseudopotential Kohn-Sham (KS) DFT, developed in our 
laboratory, taking SBKJC-type basis functions within B3LYP approximation. The triplet and singlet excitation energies 
corresponding to first singly excited configuration, are found to be in excellent agreement with TD-B3LYP calculations. Further, we 
perform the same for a set of larger molecular systems using the asymptotically corrected LC-BLYP, in addition to B3LYP. 
A systematic comparison with theoretical best estimates demonstrates the viability and suitability of current approach in 
determining optical gaps, combining predictive accuracy with moderate computational cost. 
\vspace{5mm}


\end{abstract}
\maketitle

\section{Introduction}
Over the past decades, Kohn-Sham density functional theory (KS-DFT) \citep{kohn65} has emerged as a very powerful and successful tool
for ground-state calculations of many-electron systems, such as atoms, molecules and solids \cite{becke14}. Moreover, its time-dependent 
(TD) variant, generally entitled linear-response (LR)-TDDFT has been developed for computing excitation energies \citep{runge84,casida95}, 
mostly for discrete spectrum. In comparison to the conventional wave function-based formalisms, such as configuration interaction (CI), 
multireference CI, complete active space self-consistent field, equation-of-motion coupled cluster etc., LR-TDDFT has gained popularity 
due to its reasonable trade-off between accuracy and efficiency. This is evident from its applicability to medium and large 
systems \citep{casida12}. However, despite its huge success, it has well-known difficulties regarding double excitation, charge transfer 
and Rydberg excitation. This is mainly due to the adverse effects of exchange-correlation (XC) potential within adiabatic approximation. 
A detailed analysis could be found in \citep{maitra16}. 

Besides TDDFT, attempts were also made to obtain excited states within a time-independent DFT rubric--several elegant approaches 
are available in the literature \cite{theophilou79,gross88,singh96,roy97,ayers15}. In particular, the $\Delta$SCF method \citep{ziegler77,
kowalczyk11}was developed within the standard self-consistent field iteration, by adopting non-Aufbau occupations at each iterations, and 
it targets the non-Aufbau solutions using the ground-state functional. It has favorable scaling like ground-state DFT, and hence smaller 
computational resource compared to LR-TDDFT methods. Although it already provides a good estimation of excitation energies for molecular 
systems \citep{seidu14}, its tendency to variationally collapse to the ground state is a serious concern. Such features during 
the $\Delta$SCF procedure has been characterized in recent times \citep{seidu14a}; this is particularly severe in systems with a dense 
energy spectrum near the Fermi energy levels. Several sophisticated schemes were put forth to alleviate this issue, which include a few 
constrained-DFT formalisms \citep{ziegler12, barca14, ramos16} as well as gentlest ascent dynamics and meta dynamics related methods 
\citep{li15}, etc. \color{red} These are often somehow involved, but are quite successful where TDDFT fails to perform well. Furthermore, 
a lot of recent developments have been placed in the field of unconstrained excited-state orbital relaxation that have the same complexity 
as normal ground-state or TDDFT, while using a simple $\Delta$SCF-style ansatz \citep{frank98,kowalczyk13,shea18,hait20,carter20}. \color{black}
On the other hand, orbital energies themselves have physical meaning as excitation energies \citep{van14a}, but are quite sensitive to the choice 
of density functionals and its implementations. In a recent article \citep{haiduke18}, KS-DFT has been used successfully to extract the photo-electron 
and electronic excitation spectrum for molecules from orbital energies through range-separated (RS) XC functionals. Moreover, efforts are also known  
the development of multi-reference (MR)-DFT by means of configurations mixing such as, multiconfiguration pair DFT \citep{li14} and 
MR-DFT with generalized auxiliary systems \citep{chen17}. 

In a series of articles, Becke \citep{becke16,becke18a,becke18b,becke18c}, introduced a simple novel model for estimation of lowest 
single-particle excitation energies via correlated singlet-triplet splitting (STS) energy. It 
offers an accurate, economical way to compute optical gaps in large molecules. The main focus here lies in the lowest 
singlet excited state, which are of fundamental interest in photochemistry. While triplet states can be easily obtained, \emph{in 
principle}, standard DFT cannot be used in a singlet excited state due to its multi-determinantal nature. The present scheme advocates two 
separate single-determinant DFT calculations---one for the closed-shell ground state 
and another for open-shell lowest triplet excited state, followed by a simple two-electron integral (Coulomb self energy) evaluation 
corresponding to the HOMO-LUMO transition. We note that there is no concern about standard density functionals applied on a given triplet 
excited state as it is represented by a single Slater determinant and is characterized by Fermi hole. Thus the main ingredient is the 
correlated STS energy which can be approached by means of (i) adiabatic connection theorem \citep{becke16} and (ii) virial theorem \citep{becke18a}. 
In a sense, this avoids the configuration mixing and is altogether non-empirical. 

In the present work, we have adopted the above approach for lowest single-particle excitation energy corresponding to singlet excited state 
in molecules. The required computations are performed following a pseudopotential KS-DFT implemented in Cartesian coordinate grid 
(CCG), as developed in our laboratory \citep{roy08,roy08a,roy09,roy10,roy11,ghosal16,ghosal18,mandal19,ghosal19}. 
The pertinent two-electron integral (from appropriate orbitals obtained 
after solving the lowest triplet excited state) is carried out numerically using a recently designed algorithm \citep{ghosal19}. This 
employs the Fourier convolution theorem in conjunction with a RS Coulomb interaction kernel. The latter is efficiently mapped 
onto the real grid through a simple grid optimization prescription, giving rise to some constraints in RS parameter. The 
``adiabatic" model \citep{becke16} corresponding to correlated STS energy is also provided accurately, leading to an 
easy route to compute the singlet excitation energy. Following the arguments of Becke \citep{becke18a}, an elegant ``virial" theorem is then
engaged to analyze the role of two-electron integral in determining the correlated STS energy. Both procedures are applied to calculate optical 
gaps arising from both $\pi \to \pi^{\star}$ and $n \to \pi^{\star}$ transitions in a decent number of molecules adopting B3LYP 
\citep{becke93a} functional. For a proper comparison, we carry out parallel calculations corresponding to lowest triplet 
excited states, from GAMESS quantum chemistry package \citep{schmidt93} using TD-B3LYP method, which reveals a better 
accuracy. Further, to extend the scope and applicability of this approximation, we execute the same using ``virial" theorem 
for a set of large molecular systems, and employing an additional XC functional from RS hybrid \citep{baer10} family. 
The effectiveness of this scheme is illustrated by the respective statistical analysis. The article is organized as follows. 
The underlying theorem of single-particle excitation energy along with a brief summary of our KS-DFT framework in CCG, is provided in Sec.~II. 
Section~III offers the necessary computational and technical details. Finally, the feasibility, performance and accuracy of our results are 
critically assessed in Sec.~IV. Some concluding remarks as well as the future prospects are given in Sec.~V.
  
\section{Methodology}
Let us consider an excitation of a given system, corresponding to an electronic configuration $\varphi_{i}\varphi_{f}$ from a 
closed-shell ground state. Assuming completely filled closed-shell core, this is spanned by four Slater determinants: 
$|\varphi_{i}^{\alpha} \varphi_{f}^{\alpha} \rangle$, $|\varphi_{i}^{\alpha} \varphi_{f}^{\beta} \rangle$, $|\varphi_{i}^{\beta} 
\varphi_{f}^{\alpha} \rangle$ and $|\varphi_{i}^{\beta} \varphi_{f}^{\beta} \rangle$, where $\alpha$ and $\beta$ denote up and down 
spin. Therefore, the coupled excited states are found by diagonalizing the Hamiltonian matrix in the space of above four determinants. 
As such, the singlet state is given by $|\psi_{\mathrm{S}} \rangle = \frac{1} {\sqrt{2}} \{|\varphi_{i}^{\alpha} 
\varphi_{f}^{\beta}\rangle-|\varphi_{i}^{\beta} \varphi_{f}^{\alpha}\rangle\}$, whereas the three degenerate triplet states are 
written as follows: $\psi_{\mathrm{T}}=|\varphi_{i}^{\alpha} \varphi_{f}^{\alpha} \rangle \quad \mathrm{or} \quad \frac{1}{\sqrt{2}} 
\{|\varphi_{i}^{\alpha} \varphi_{f}^{\beta}\rangle+ |\varphi_{i}^{\beta} \varphi_{f}^{\alpha}\rangle\} \quad \mathrm{or} 
\quad|\varphi_{i}^{\beta} \varphi_{f}^{\beta} \rangle$. The corresponding singlet and triplet energies (identified by ``S" and ``T" 
subscripts) are,  
\begin{equation}
E_{\mathrm{S}}=E^{\alpha\beta}+K_{if}
\end{equation}
and 
\begin{equation}
E_{\mathrm{T}}=E^{\alpha\alpha} \quad \mathrm{or} \quad E^{\alpha\beta}-K_{if} \quad \mathrm{or} \quad E^{\beta\beta},
\end{equation}
where $E^{\sigma_1\sigma_2}$ is the energy of a given determinant of form $|\varphi_{i}^{\sigma_1}\varphi_{f}^{\sigma_2}\rangle$, 
$(\sigma_1, \sigma_2) \! \in \!\{\alpha, \beta\}$, and $K_{if}$ is the 2e$^-$ integral (or Coulomb self-energy of product of 
transition orbitals) defined as, 
\begin{equation}
K_{if}=\int\int\frac{\varphi_i(\rvec_1)\varphi_f(\rvec_1)\varphi_i(\rvec_2)\varphi_f(\rvec_2)}{|\rvec_1-\rvec_2|} \ d\rvec_1 d\rvec_2.
\end{equation}
Now combining Eqs.~(1), (2), one can connect singlet and triplet excitation energy as follows, 
\begin{equation}
E_{0\mathrm{S}}=E_{0\mathrm{T}}+2K_{if}, 
\end{equation}
where $E_{0\mathrm{S}}=E_{\mathrm{S}}-E_0, \ E_{0\mathrm{T}}=E_{\mathrm{T}}-E_0$, while $E_0$ signifies the ground-state energy of 
a certain closed-shell system. But Eq.~(4) is highly inaccurate for singlet excitation energy, $E_{0\mathrm{S}}$. The problem lies in the 
determination of $2K_{if}$ term, also called the zeroth-order (or uncorrelated) STS energy. In order to tackle this 
issue, recently some novel proposals (semi-empirical \citep{becke16} as well as non-empirical \citep{becke16,becke18a})
have appeared in the literature to derive a simplified formula for correlated STS energy. 

\subsection{The adiabatic connection theorem}
One can make use of the well-known ``adiabatic connection" theorem \citep{harris74} to approach the correlated STS energy in 
single-particle excitations, which we are discussed here. For a specific excited configuration, all the orbitals involved in 
Eqs.~(1), (2) are same. In that case, singlet and triplet states have same density and non-interacting kinetic energy. Then 
their energy difference can be written as ($\Delta E_{\mathrm{STS}}^{0} = 2K_{if}$), 
\begin{eqnarray}
\Delta E_{\mathrm{STS}}=\Delta E_{\mathrm{STS}}^{0}+\Delta E_{\mathrm{ST}}^{\mathrm{corr}}, \nonumber \\
\Delta E_{\mathrm{STS}}=2K_{if}+\Delta E_{\mathrm{ST}}^{\mathrm{corr}},
\end{eqnarray}
where $\Delta E_{\mathrm{ST}}^{\mathrm{corr}}$ represents the singlet-triplet correlation energy difference. Recently,  
a non-empirical formula \cite{becke16} has been proposed for it, based on the inter-electronic cusp condition, and its effect 
on electron correlation. Accordingly, it can be expressed as, 
\begin{equation}
\Delta E_{\mathrm{ST}}^{\mathrm{corr}}=-0.4 \ \int 4 \ \varphi_{i}^{2}(\rvec_1)\varphi_{j}^{2} (\rvec_1) \  z_{\mathrm{C}}^{2} 
\bigg[1-\frac{ \ln(1+z_{\mathrm{C}})} {z_{\mathrm{C}}}\bigg]d\rvec_1.
\end{equation}
The only unknown quantity, the correlation length $z_{\mathrm{C}}$, measures the spatial extent of electron correlation in 
configuration, $\varphi_i \varphi_f$. Now if one allows a ``strictly correlated electrons" limit, $z_{\mathrm{C}}$ can be 
written in terms of two-electron integral, $K_{if}$, as in the following,   
\begin{equation}
0.4 z_{\mathrm{C}}^{2}\int \ 4 \ \varphi_{i}^{2}(\rvec_1)\varphi_{j}^{2}(\rvec_1) \ d\rvec_1=2K_{if}.
\end{equation}
A detailed derivation could be found in \citep{becke16}. 

\subsection{The virial exciton model}
From the above discussion, it is clear that, one may add a correlation correction term, $\Delta E_{\mathrm{ST}}^{\mathrm{corr}}$, 
to $2K_{if}$ to recover the desired correlated STS energy. As suggested in \cite{becke18a}, one may also further proceed to 
approximate this by noting that it comprises the kinetic and potential energy contributions, 
\begin{equation}
\Delta E_{\mathrm{STS}}^{\mathrm{corr}} = \Delta T_{\mathrm{STS}}^{\mathrm{corr}}+\Delta V_{\mathrm{STS}}^{\mathrm{corr}}.
\end{equation}
Now invoking the standard virial theorem, the above equation can be simplified as, 
\begin{equation}
\Delta E_{\mathrm{STS}}^{\mathrm{corr}} = \frac{1}{2}\Delta V_{\mathrm{STS}}^{\mathrm{corr}}.
\end{equation}
One may further presume that the non-local part of $\Delta V_{\mathrm{STS}}^{\mathrm{corr}}$ is dominated by pair-density effects. 
In that scenario, correlation would then lower the potential energy of singlet state relative to triplet state, by 
$\Delta V_{\mathrm{STS}}^{\mathrm{corr}}=-2K_{if}$. Hence, one can write,  
\begin{equation}
\Delta E_{\mathrm{STS}}^{\mathrm{corr}}=-K_{if}. 
\end{equation}
which surprisingly leads to a very simple relation as below,  
\begin{equation}
\Delta E_{\mathrm{STS}}=2K_{if}-K_{if}=K_{if}.
\end{equation}
It may be noted that, this model (as well as that in II(A)) is also a purely non-empirical one. But obviously, this
expression for correlated STS energy is much simpler involving only the two-electron integral. 

\subsection{The semi-empirical approach}
As implied, both the routes in Eqs.~(5), (11) are purely non-empirical, suggesting $\Delta E_{\mathrm{STS}}$ to be less than 
$2K_{if}$. This prompts one to formally define a molecule-independent re-scaling parameter $f$ such that, 
\begin{equation}
\Delta E_{\mathrm{STS}}=2fK_{if}, \quad 0<f<1.
\end{equation}
and if determined semi-empirically, it might offer good excitation energies overall, as long as de-localization error \citep{perdew82} 
is not a serious issue. This led to a semi-empirical technique \citep{becke16} to obtain an optimum value of $f$. This was accomplished
in \cite{becke16} by fitting the results with best estimated theoretical excitation energy data set of \citep{silva10}, 
resulting in a value for $f$ as 0.486. It is surprisingly close to 0.5, as also appears from a consideration of II(B). The same 
author has also reported that the obtained mean average error (using this $f$) remains similar to that in non-empirical 
calculations, in II(A). 

\section{Computational details}
Our desired quantity, the singlet excitation energy $E_{0\mathrm{S}}$ can be obtained from two single-determinant calculations 
(both are spin polarized), namely, one for closed-shell ground state and another for fully relaxed triplet state. These are 
computed using our in-house pseudopotential KS-DFT program in CCG \cite{indft19}. An initial version was first 
developed by the corresponding author in 2008, on the basis of works \citep{roy08,roy08a,roy09,roy10,roy11}, which has been later 
extended by his group \cite{ghosal16,ghosal18,mandal19,ghosal19}. Since the details have been published in the above references, 
here we give only some essential aspects. Accordingly, the single-particle KS equation in presence of pseudopotential is written as 
(atomic unit employed unless stated otherwise), 
\begin{equation}
\bigg[ -\frac{1}{2} \nabla^2 + v^{\mathrm{p}}_{\mathrm{ion}}(\rvec) + v_{\mathrm{ext}}(\rvec) + v_{\mathrm{h}}[\rho(\rvec)] 
+ v_{\mathrm{xc}}[\rho(\rvec)] \bigg] \varphi_i^{\sigma}(\rvec) = \epsilon_i \varphi_i^{\sigma}(\rvec),
\end{equation}
where $v^{\mathrm{p}}_{\mathrm{ion}}$ denotes the ionic pseudopotential, written as below, 
\begin{equation}
v^{\mathrm{p}}_{\mathrm{ion}}(\rvec) = \sum_{\Rvec_a} v^{\mathrm{p}}_{\mathrm{ion},\mathrm{a}} (\rvec-\Rvec_a).
\end{equation}
In the above equation, $v^{\mathrm{p}}_{\mathrm{ion},\mathrm{a}}$ signifies ion-core pseudopotential associated with atom A, 
situated at $\Rvec_a$. The classical Coulomb (Hartree) term, $v_{\mathrm{h}}[\rho(\rvec)]$ describes usual electrostatic 
interaction amongst valence electrons, whereas $v_{\mathrm{xc}}[\rho(\rvec)]$ represents the non-classical XC part of latter, 
and $\{ \varphi^{\sigma}_{i},\sigma= \alpha \quad \mathrm{or} \quad  \beta \}$ corresponds to a set of $N$ occupied orthonormal 
spin molecular orbitals (spin-MOs).  
 
Now various quantities like localized basis function, two-electron potential, MOs and electron density are directly 
set up on a 3D cubic box, 
\begin{eqnarray}
r_{i}=r_{0}+(i-1)h_{r}, \quad i=1,2,3,....,N_{r}~, \quad r_{0}=-\frac{N_{r}h_{r}}{2}, \quad  r \in \{ x,y,z \},
\end{eqnarray}
where $h_{r}, N_r$ denote grid spacing and total number of points along three directions respectively. While one-electron 
contributions of KS-Fock matrix are evaluated through well-established recursion relations, the two-electron matrix elements 
are computed via direct numerical integration in the grid, 
\begin{equation}
F_{\mu \nu}^{\mathrm{hxc}} = \langle \chi_{\mu}(\rvec_g)|v_{\mathrm{hxc}}(\rvec_g)|\chi_{\nu}(\rvec_g) \rangle = h_x h_y h_z 
\sum_g \chi_{\mu}(\rvec_g) v_{\mathrm{hxc}}(\rvec_g) \chi_{\nu}(\rvec_g).
 \end{equation}
where $v_{hxc}$ refers to the classical (Hartree) and non-classical (XC) potential combined. The construction of HF exchange 
in KS-Fock 
matrix has been documented in our earlier work \citep{ghosal19}; hence not repeated here. 

\color{red}
The actual implementation of the central quantity, $K_{if}$, is quite different from the original prescription in 
\citep{becke16}. Here, we use the Fourier convolution theorem (FCT) using a range-separated technique in Coulomb interaction 
kernel, instead of the multi-center numerical integral procedure put forth in \citep{becke88c}. According to 
\citep{ghosal19}, this two-electron integral $K_{if}$ can be rewritten as,  
\begin{eqnarray}
K_{if}=\int\int\frac{\varphi_i(\rvec_1)\varphi_f(\rvec_1)\varphi_i(\rvec_2)\varphi_f(\rvec_2)}{|\rvec_1-\rvec_2|} \ d\rvec_1 
d\rvec_2 \nonumber \\
=\int \varphi_i(\rvec_1) \varphi_f(\rvec_1) v_{if}(\rvec_1) d\rvec_1.  
\end{eqnarray}
Now, the main concern lies in the evaluation of $v_{if}$ integral which is related to electro static potential integral 
evaluation in \citep{ghosal19}. Again, this can be recast as,
\begin{eqnarray}
v_{if}(\rvec_1)=\int\frac{\varphi_i(\rvec_2) \varphi_f(\rvec_2)}{|\rvec_1 - \rvec_2|} d\rvec_2 \nonumber \\
= \int \frac{\varphi_{if}(\rvec_2)}{|\rvec_1 - \rvec_2|} d\rvec_2 = \varphi_{if}(\rvec_1) \star v^{\mathrm{c}}(\rvec_1)
\end{eqnarray} 
The last expression is in terms of convolution integral, where $\varphi_{if}$ denotes simple multiplication of ith and fth MO 
from lowest triplet excited state and $v^{\mathrm{c}}(\rvec)$ represents the Coulomb interaction kernel. Now one can invoke 
the FCT to further  simplify this integral,
\begin{equation}
v_{if}(\rvec) = \mathcal{F}^{-1}\{v^{\mathrm{c}}(\kvec)\varphi_{if}(\kvec)\} \quad \mathrm{where} \quad \varphi_{if}(\kvec)=
\mathcal{F}\{ \varphi_{if}(\rvec)\}
\end{equation}
Here $v^{c}(\kvec)$ and $\varphi_{if}(\kvec)$ stand for Fourier integrals of Coulomb kernel and MOs respectively. The main concern 
lies in an accurate mapping of the former, which has singularity at $\rvec \to 0$. In order to alleviate this problem, we apply 
a simple RS technique from \citep{ghosal19}, expanding the kernel into long- and short-range components with a suitably chosen 
RS parameter ($\zeta$). This gives rise to the following expression,  
\begin{eqnarray}
v^{c}(\rvec)=\frac{\mathrm{erf}(\zeta\rvec)}{\rvec}+\frac{\mathrm{erfc}(\zeta\rvec)}{\rvec} \nonumber \\
v^{c}(\rvec_g)= v^{c}_{long}(\rvec_g) + v^{c}_{short}(\rvec_g). 
\end{eqnarray}
In the above equation, $\mathrm{erf}(x)$ and $\mathrm{erfc}(x)$ denote error function and its complement respectively, while 
the second expression is written in real-space CCG. The Fourier integral of Coulomb kernel can be separated out as follows: 
(i) FT of short-range part is treated analytically and (ii) long-range portion is computed directly from FFT of corresponding 
real-space CCG values. Then the remaining problem lies in finding an optimum value of parameter $\zeta_{\mathrm{opt}}$ for 
successful mapping of Coulomb kernel in CCG from \emph{first principles}. In this regard, we proposed a simple procedure which is 
found to be sufficiently accurate for lowest triplet excited states (as exemplified in results that follows) as well. This prompts us to write, 
\begin{equation}
\zeta_{\mathrm{opt}} \equiv \underset{\zeta }{\mathrm{opt}} \ 
E_{\mathrm{tot}} = \underset{N_x,N_y,N_z} {\mathrm{opt}} E_{\emph{tot}}, \quad \mathrm{at \ fixed} \ h_{\rvec},
\end{equation}
using a suitably defined constraint \cite{ghosal19} $(\zeta \times L =7 )$, where $L\ (= N_\rvec h_\rvec; \ \rvec \in 
\{x,y,z\})$ 
is the smallest side of simulating box. In the same spirit, other necessary quantities such 
as correlation length, $z_{\mathrm{C}}$ and singlet-triplet correlation energy difference, $\Delta E_{\mathrm{C}}$ are directly 
computed in real-space grid using pseudo KS orbitals $\varphi_i$ and $\varphi_f$. 
 
\color{black}
The strategy as outlined for HF exchange component, is implemented in case of B3LYP, containing a fixed amount of former with conventional 
DFT XC functional. As such, the XC energy corresponding to this functional is defined as follows, 
\begin{equation}
E_{\mathrm{xc}}^{\mathrm{B3LYP}} = (1-a_0)E^{\mathrm{x}}_{\mathrm{LSDA}}+a_0 E^{\mathrm{x}}_{\mathrm{HF}}+ 
a_{\mathrm{x}} E^{\mathrm{x}}_{\mathrm{B88}}+a_{\mathrm{c}} E^{\mathrm{c}}_{\mathrm{LYP}} + (1-a_{\mathrm{c}})
E^{\mathrm{c}}_{\mathrm{VWN}}.
\end{equation}
Throughout our current presentation, we use the recommended values of $a_0, a_{\mathrm{x}}, a_{\mathrm{c}}$, as advocated in 
\cite{stephens94}, i.e., $0.2$, $0.72$, $0.81$ for B3LYP. This is computed using KS orbitals 
obtained from the solution of Eq.~(13) in real-space CCG. The relevant LDA- and GGA-type functionals 
in connection with B3LYP are: (i) Vosko-Wilk-Nusair (VWN)--with the homogeneous electron gas correlation proposed in 
parametrization formula V \citep{vosko80} (ii) B88--incorporating Becke \citep{becke88a} semi-local exchange (iii) Lee-Yang-Parr (LYP) 
\citep{lee88} semi-local correlation. All these are adopted from density functional repository program \citep{repository} except LDA.  
 
The present work employs SBKJC \citep{stevens84} effective core potential basis sets for species containing Group II elements. 
These are imported from EMSL Basis Set Library \citep{feller96}. The norm-conserving pseudopotential matrix elements in contracted 
basis are collected from GAMESS \citep{schmidt93}. \color{red}The triplet calculations refer to \emph{restricted} open-shell, so that 
orbitals $\varphi_i$ and $\varphi_f$ are well defined. The convergence criteria 
imposed in this communication are slightly tighter than our earlier works \citep{roy08, roy08a, roy11}; this is to generate 
accurate orbital energies, especially for triplet excited states. \color{black}Changes in following quantities were followed during 
self-consistent field process, \emph{viz.,} (i) orbital energy difference between two successive iterations and (ii) absolute 
deviation in density matrix elements. They both were required to remain below a prescribed threshold set to $10^{-8}$ a.u.; this 
ensured that total energy maintained a convergence of at least this much, in general. In order to perform discrete Fourier transform, 
standard FFTW3 package \cite{fftw05} was invoked. The resulting generalized matrix-eigenvalue problem is solved through standard 
LAPACK routine \cite{anderson99} easily. All molecular calculations are performed with calculated geometry 
(B3LYP XC functional and cc-PVTZ basis set), taken from NIST database \citep{johnson16} (otherwise stated below). Other details 
including scaling properties may be found in references \citep{roy08,roy08a,roy09,roy10,roy11, ghosal16, ghosal18, mandal19, ghosal19}.

\section{Result and Discussion}
A set of 10 selective organic chromophores ($\pi$ containing molecules) from our previous work \citep{ghosal19} has been chosen, 
excluding those which have degenerate frontier orbitals, as the method in its present form is not applicable for these. These results 
pertain to the pseudopotential approximation neglecting the effects of core electrons. This may, by and large, be justifiable on the  
ground that only valence excitations are taken into account. By doing so, one can make a balance between accuracy and cost. To 
establish the effectiveness of this approach, we have chosen ``SBKJC" type of pseudopotential basis set which does not contain any diffuse 
function, and this suffices in our present scenario. It may be emphasized that, in this exploratory study, only lowest singlet and 
triplet excited state corresponding to first single excitation, is pursued for each molecule. 

Each calculation begins by optimizing the non-uniform grid for a given molecule, both in its closed-shell ground and lowest excited 
triplet state separately, employing a simple procedure 
as reported earlier \citep{ghosal16, ghosal18}. Individual singlet excitation energies (in eV) through B3LYP XC functional, are
displayed in Table~I, along with mean absolute error (MAE) and mean error (ME) statistics. It contains both non-empirical and 
empirical excitation values as following: (i) PR$_1$ represents Eq.~(12) using semi-empirical re-scaling parameter $f=0.486$ (ii) 
PR$_2$ presents Eq.~(5) using non-empirical model from ``adiabatic connection" (iii) PR$_3$ refers to results from Eq.~(11) using 
non-empirical model from virial theorem. To put things in perspective, column 6 reports the respective TD-B3LYP energies employing 
same functional and basis set, computed from GAMESS\cite{schmidt93}. The correlation lengths $z_\mathrm{C}$ manifestly fall 
into two distinct 
groups with $\pi \to \pi^{\star}$ and $n \to \pi^{\star}$ transition--a feature that has been observed in \citep{becke16}. 
\color{red}Furthermore, 
in consonance with their finding, we also witness a similar kind of pattern in the calculated $z_{\mathrm{C}}$ values, having smaller 
$\varphi_i \varphi_f$ overlap for the $n \to \pi^{\star}$ transition for current set of molecules. \color{black}
Overall, the obtained singlet excitation energies are within 
fraction of eV compared to the TD-B3LYP results. As expected from \citep{becke18a}, the correlated virial theorem (PR$_3$) confers 
remarkable improvement over the adiabatic connection formula (PR$_2$). Here, the MAE value is about two times smaller compared to 
the latter one. On the other hand, the semi-empirical approach (PR$_2$) performs very close to PR$_3$ results, as evidenced
from the proximity in their corresponding MAE values. The worst case candidate is propene, with an excitation energy $0.55$ eV 
too low, in comparison to TD-B3LYP result. We note that the MAE and ME remain same in magnitude and sign, indicating a systematic 
error in calculated values. Out of three methods, the virial PR$_3$ route provides the best estimate in exciting energies; quite
competitive with that of TD-B3LYP, with an MAE of 0.34 eV. 

\begingroup
\squeezetable
\begin{table}               
\caption{\label{tab:table1} Singlet excitation energies (in eV) using B3LYP functional. See text for details.}
\begin{ruledtabular}
\begin{tabular} {lcccccc}
Molecule & State & PR$_1$ & PR$_2$ & PR$_3$ & TD-B3LYP \citep{schmidt93} & $z_\mathrm{C}$  \\ 
\cline{1-7}
Ethylene & B$_{1u}$($\pi \rightarrow \pi^{\star}$) & 7.78 & 7.63 & 7.87 & 8.09 & 2.97 \\
Propene & A$^{\prime}$($\pi \rightarrow \pi^{\star}$) & 7.18 & 7.05 & 7.26 & 7.81 & 2.99 \\
1,3-Butadiene (E) & B($\pi \rightarrow \pi^{\star}$) & 5.63 & 5.42 & 5.70 & 6.02 & 3.29 \\
1,3-Pentadiene (E) & A$^{\prime}$($\pi \rightarrow \pi^{\star}$) & 5.47 & 5.28 & 5.54 & 5.88 & 3.29 \\
1,3,5-Hexatriene (E) & B$_{u}$($\pi \rightarrow \pi^{\star}$) & 4.38 & 4.14 & 4.44 & 4.79 & 3.53 \\
2,4-Hexadiene (E,E) & B$_{u}$($\pi \rightarrow \pi^{\star}$) & 5.39 & 5.20 & 5.45 & 5.79 & 3.27 \\
1,3-Cyclo-pentadiene & A$^{\prime}$($\pi \rightarrow \pi^{\star}$) & 5.12 & 5.03 & 5.17 & 5.28 & 2.98\\
Thiophene & B$_{2}$($\pi \rightarrow \pi^{\star}$) & 5.61 & 5.31 & 5.66 & 6.02 & 3.99 \\
Formaldehyde & A$_{2}$($n \rightarrow \pi^{\star}$ & 3.52 & 3.63 & 3.53 & 3.98 & 1.30 \\
Acetaldehyde & A$^{\prime \prime}$($n \rightarrow \pi^{\star}$) & 4.67 & 4.75 & 4.68 & 5.07 & 1.44 \\
\hline
MAE          & -- & 0.40 & 0.53 & 0.34 & -- & --\\
ME          & -- & 0.40 & 0.53 & 0.34 & -- & --\\
\end{tabular}
\end{ruledtabular}
\end{table}
\endgroup

Next we analyze the effects of triplet excitations and correlated STS term separately for B3LYP functional in Table~II, 
giving a side-by-side comparison with TD-B3LYP in both cases. As singlet excitation energies are obtained from these two 
above quantities, a detailed assessment of them is worthwhile considering. Due to the superior performance, henceforth in 
future proceedings, we engage only the PR$_3$. The $E_{0\mathrm{T}}$ error metrics are: MAE = 0.40 eV 
and ME = $-$0.39 eV, having very similar magnitude, but different sign. The triplet MAE remains only marginally larger than singlet MAE 
values of PR$_3$ results of previous table. On the other hand, STS measures read: MAE = 0.70 eV and ME = 0.69 eV, showing very close 
agreement in magnitude and also importantly without changing sign. However magnitude of these errors (MAE and ME) 
in STS are relatively larger than the $E_{0T}$ errors. A careful look on respective ME in these two quantities reveals that, 
ME in STS is about 75\% larger compared to that in E$_{0T}$, but with an opposite sign. And that makes the singlet excitation energies more 
close to TD-B3LYP results, occurring through a systematic error cancellation. This leads us to conclude that, \emph{STS term and not}
$E_{0\mathrm{T}}$, is the major source of error. It is worthwhile mentioning that, the success of this approach relies on an accurate 
estimation of the two-electron integral, which in turn depends on the accuracy of triplet excited state. Thus it is evident that our current 
prescription can be reliably used for determining optical gaps using the non-empirical time-independent methods 
of Sec.~II, at least for small molecular systems. 

\begingroup
\squeezetable
\begin{table}               
\caption{\label{tab:table2} Triplet excitation energies and correlated STS energies (in eV) using B3LYP functional.}
\begin{ruledtabular}
\begin{tabular} {lccccc}
Molecule & State & \multicolumn{2}{c}{$E_{0\mathrm{T}}$} & \multicolumn{2}{c}{$\Delta E_{\mathrm{STS}}$} \\
\cline{3-4} \cline{5-6} 
 &  & Ref.~\citep{indft19} & TD-B3LYP & PR$_3$ & TD-B3LYP \\ 
\cline{1-6}
Ethylene & B$_{1u}$($\pi \rightarrow \pi^{\star}$) & 4.47 & 4.03 & 3.40 & 4.06 \\
Propene  & A$^{\prime}$($\pi \rightarrow \pi^{\star}$) & 4.44 & 4.03 & 2.82 & 3.78 \\
1,3-Butadiene (E) & B($\pi \rightarrow \pi^{\star}$) & 3.26 & 2.71 & 2.44 & 3.31 \\
1,3-Pentadiene (E) & A$^{\prime}$($\pi \rightarrow \pi^{\star}$) & 3.24 & 2.71 & 2.30 & 3.17 \\
1,3,5-Hexatriene (E) & B$_{u}$($\pi \rightarrow \pi^{\star}$)  & 2.42 & 1.85 & 2.02 & 2.94 \\
2,4-Hexadiene (E,E) & B$_{u}$($\pi \rightarrow \pi^{\star}$) & 3.22 & 2.70 & 2.23 & 3.09 \\
1,3-Cyclo-pentadiene & A$^{\prime}$($\pi \rightarrow \pi^{\star}$) & 3.21 & 2.70 & 1.96 & 2.58 \\
Thiophene & B$_{2}$($\pi \rightarrow \pi^{\star}$) & 3.88 & 3.47 & 1.78 & 2.55 \\
Formaldehyde & A$_{2}$($n \rightarrow \pi^{\star}$) & 3.19 & 3.20 & 0.34 & 0.78 \\
Acetaldehyde & A$^{\prime \prime}$($n \rightarrow \pi^{\star}$) & 4.39 & 4.44 & 0.29 & 0.24 \\
\hline
MAE          & -- & 0.40 & -- & 0.70 & -- \\
ME          & -- & $-$0.39 & -- & 0.69 & --  \\
\end{tabular}
\end{ruledtabular}
\end{table}
\endgroup
 
\begingroup
\squeezetable
\begin{table}               
\caption{\label{tab:table3} Excitation energies (in eV) of organic chromophores from ``virial theorem".}
\begin{ruledtabular}
\begin{tabular} {lc|cc|ccc}
Molecule & State & \multicolumn{2}{c}{$E_{0\mathrm{T}}$} & \multicolumn{2}{c}{$E_{0\mathrm{S}}$ (PR$_3$) } & Lit.$^\ddagger$ \\
\hline
 &  & B3LYP & LC-BLYP & B3LYP & LC-BLYP &    \\ 
\hline
Cyclopropene & B$_{2}$($\pi \rightarrow \pi^{\star}$) & 4.03 & 4.05 & 7.04 & 7.07 &  7.01 \\
Norbornadiene & A$_{2}$($\pi \rightarrow \pi^{\star}$) & 4.62 & 4.23 & 5.77 & 5.45 &  4.91 \\
Naphthalene & B$_{2u}$($\pi \rightarrow \pi^{\star}$) & 3.22 & 3.53 & 4.74 & 5.20 &  4.64 \\
Furan & B$_{2}$($\pi \rightarrow \pi^{\star}$) & 4.15 & 4.25 & 6.65 & 6.85 & 6.57 \\
Pyrrole & B$_{2}$($\pi \rightarrow \pi^{\star}$) & 4.45 & 4.52 & 6.88 & 7.07 & 6.85 \\
Pyridine & B$_{1}$($n \rightarrow \pi^{\star}$)  & NC & 4.22 & NC & 4.65  & 4.74 \\
Pyridazine & B$_{3u}$($n \rightarrow \pi^{\star}$) & 2.76 & 2.78 & 3.35 & 3.34 & 3.57 \\
s-tetrazine & B$_{2}$($\pi \rightarrow \pi^{\star}$) & 1.39 & 1.63 & 1.83 & 2.07 & 2.15 \\
Acetone & A$_{2}$($n \rightarrow \pi^{\star}$) & 3.71 & 3.80 & 3.98 & 4.08 & 4.11 \\
Formamide & A$^{\prime \prime}$($n \rightarrow \pi^{\star}$) & NC & 5.03 & NC & 5.25 & 5.42 \\
Acetamide & A$^{\prime \prime}$($n \rightarrow \pi^{\star}$) & 5.20 & 5.15 & 5.45 & 5.37 & 5.46 \\
Propanamide & A$^{\prime \prime}$($n \rightarrow \pi^{\star}$) & 5.64 & 5.19 & 6.07 & 5.40 & 5.49 \\
Thymine & A$^{\prime}$($\pi \rightarrow \pi^{\star}$) & 3.51 & 3.59 & 5.50 & 5.73 & 5.48 \\
Uracil & A$^{\prime \prime}$($n \rightarrow \pi^{\star}$) & 4.28 & 3.76 & 5.66 & 5.86 & 4.66 \\
Imidazole & A$^{\prime}$($\Pi \rightarrow \pi^{\star}$) & NC & 4.75 & NC & 7.15 & 6.89 \\
E-Octatetraene & B$_{u}$($\pi \rightarrow \pi^{\star}$) & 1.93 & 1.92 & 4.32 & 3.69 & 4.22 \\
\hline
MAE from Lit. results\citep{becke18a} . &   &  & & 0.27 & 0.23    &  \\
ME from Lit. results \citep{becke18a}. &   &  & & $-$0.16 & $-$0.07    &   \\
\hline
\color{red}
MAE from TBE-2 results \citep{silva10} . &   &  & & \color{red}0.38 & \color{red}0.45    &  \\
\color{red}ME from TBE-2 results \citep{silva10}. &   &  & & \color{red}$-$0.05 & \color{red}$-$0.09    &   \\
\end{tabular}
\end{ruledtabular}
\begin{tabbing}
\color{red}
$^a$MAE and ME values from RO-PBE0 calculations using 6-31G$^{\star}$ are 0.37 and $-$0.22 
respectively \citep{kowalczyk13}. \\ 
\color{red}
$^b$MAE and ME values from RO-LC$\omega$PBE0 calculations using 6-31G$^{\star}$ are 0.31 and $-$0.05 
respectively \citep{kowalczyk13}. \\ 
\color{black}
$^{\dagger}$NC denotes ``not converged". \hfill \hspace{200pt} $^{\ddagger}$ This corresponds to $E_{0\mathrm{S}}$, from \citep{becke18a}.
\end{tabbing}
\end{table}
\endgroup

At this stage, in order to assess the validity and usefulness, an additional set of larger molecular systems are taken up in 
Tables~III and IV. We consider a representative set of $16$ organic 
chromophores and $5$ linear acene molecules with 2--6 rings for which accurate theoretical reference data is available \citep{junior10}. 
All the molecular geometry of Table~III are taken from \citep{junior10} whereas for Table~IV, these are optimized with B3LYP and cc-pVDZ 
basis set using GAMESS. Since the present scheme has earlier been verified to deliver practically identical results as that of GAMESS, 
for brevity, we have employed the same for calculation of $E_{0}$ and $E_T$ in the same basis. In addition to B3LYP, another XC functional 
from RS hybrid family, has been invoked. Accordingly, the inter-electronic space is separated through the use of a 
RS-operator $g(\gamma,\rvec)$ such that, 
\begin{equation}
\frac{1}{\rvec}=\frac{\tilde{g}(\gamma,\rvec)}{\rvec}+\frac{g(\gamma,\rvec)}{\rvec}, 
\end{equation}
where $\tilde{g}(\gamma,\rvec)$ represents the complementary RS-operator. Here, $\gamma$ is an important RS parameter which 
has a pivotal role to adjust the contribution of HF exchange between short-range (SR) and long-range (LR) region, for a specific 
$g(\gamma,\rvec)$. In particular, we consider the long-range correction (LC) scheme presented in \citep{iikura01}, given by,  
\begin{equation}
\begin{rcases}
E_{\mathrm{xc}}^{\mathrm{LC}} = E^{\mathrm{dfa,sr}}(\gamma)+E^{\mathrm{x,lr}}(\gamma)+E^{\mathrm{c,dfa}}, \\
g(\gamma,\rvec)=\mathrm{erf}(\gamma\rvec) \quad \mathrm{and} \quad \tilde{g}(\gamma,\rvec)=\mathrm{erfc}(\gamma,\rvec). 
\end{rcases}
\end{equation}
It uses full HF exchange at LR region ($E^{\mathrm{x,lr}}$) in contrast to the conventional hybrid functionals, and successfully 
mitigates the lack of having correct asymptotic behaviour of exact exchange at LR region present in hybrid family. But at the same time, it fails 
to achieve the standard energy calculations at the level of hybrid ladder. To be consistent with B3LYP, we utilize the particular LC-BLYP (with 
$\gamma=0.33$), where interelectronic distance dependent B88 exchange \citep{iikura01} (in place of original B88 exchange \citep{becke88a}) and 
LYP \citep{lee88} semi-local correlational functionals were used as $E^{\mathrm{dfa,sr}}(\gamma)$ and $E^{\mathrm{c,dfa}}$ respectively.

\begingroup
\squeezetable
\begin{table}               
\caption{\label{tab:table4} Excitation energies (in eV) of linear acenes from ``virial theorem".}
\begin{ruledtabular}
\begin{tabular} {lcc|ccc}
Number of rings & \multicolumn{2}{c}{$E_{0\mathrm{T}}$} & \multicolumn{2}{c}{$E_{0\mathrm{S}}$ (PR$_3$)} & Lit.$^\dagger$  \\
\hline
 & B3LYP & LC-BLYP & B3LYP & LC-BLYP & \\ 
\hline
2 & 3.24 & 3.56 & 4.75 & 5.22 & 4.65 \\
3 & 2.22 & 2.46 & 3.71 & 4.43 & 3.58 \\
4 & 1.55 & 1.81 & 2.85 & 3.41 & 2.75 \\
5 & 1.08 & 1.32 & 2.30 & 2.96 & 2.22 \\
6 & 0.75 & 0.98 & 1.89 & 2.50 & 1.82 \\
\hline
MAE from Lit. results \citep{becke18a}& & & 0.09  &  0.70 &    \\
ME from Lit. results \citep{becke18a}& & & $-$0.10  &  $-$0.70 &     \\
\hline
\color{red}MAE from experimental results \citep{grimme03}& & & \color{red}0.09  &  \color{red}0.60 &    \\
\color{red}ME from experimental results \citep{grimme03}& & & \color{red}$-$0.16  &  \color{red}$-$0.60 &     \\
\end{tabular}
\end{ruledtabular}
\begin{tabbing}
$^{\dagger}$ This corresponds to $E_{0\mathrm{S}}$, from \citep{becke18a}.
\end{tabbing}
\end{table}
\endgroup

The calculated $E_{0\mathrm{S}}$, as computed above, are compared with the all-electron results from Becke's ``virial theorem" method 
\citep{becke18a} along with respective MAE and ME values for two concerned functionals at the bottom. As in Table~II, here also, due 
to the proximity of PR$_3$ results to that of TD-B3LYP, it suffices to report only PR$_3$ excitation energies. 
For sake of completeness, this also accompanies the lowest triplet excitation energies. 
Note that, for three molecules (pyridine, formamide, imidazole), the spin-restricted triplet calculations did not lead to convergence 
in the default options, for B3LYP. Hence they are not included in MAE and ME evaluations.  
It is evident that the performance of B3LYP (excitation energies are overestimated) is usually much more consistent than that of LC-BLYP. 
A systematic error cancellation may be involved in these 
calculation. On the other hand, the MAE value of LC-BLYP (0.23) is slightly 
lower than that of B3LYP (0.27), but did not outperform extensively as was expected from the foregoing discussion. Initially, one may 
draw the conclusion that the effect of full HF exchange at long-range does not have significant impact on excitation energies, despite 
the fact that it enriches the behavior of frontier orbitals involved in $K_{if}$ calculations. Such discrepancy has occurred due to the 
fact that we assume $\gamma$ is independent of system size. However, it has been shown that it is possible to attain a higher level of 
performance provided one treats $\gamma$ as a system-dependent parameter (functional of $\rho$) tuned from \emph{first principles} \citep{baer10}. 
It is our belief that, an optimally tuned $\gamma$ (perhaps in the spirit of size dependency principle) will provide better results 
than the conventional hybrid and RS hybrid functionals. Note that, the less dramatic performance of LC-BLYP in both ground and triplet excited state 
has negligible effect on singlet excitation energy, relative to the STS term, which is in harmony with the observation made in 
\citep{becke18b}. 
\color{red}
Additionally, we have also estimated MAE and ME values from TBE-2 results \citep{silva10}. It is found that 
these in Table III (with respect to TBE-2), do not alter significantly from their counterparts in 
Becke \citep{becke18a}, which was of course, expected. So the basic conclusions will remain same. 
Furthermore, we have also quoted restricted open-shell KS (ROKS) error statistics (MAE and ME) for different XC functionals in footnote of 
Table~III from \citep{kowalczyk13}. They are all-electron ROKS lowest vertical excitation results computed using 6-31G$^{\star}$ basis set. Our 
performance (MAE and ME from TBE-2) are in harmony with ROKS MAE and ME values. \color{black}
A similar qualitative trend is also found for linear acenes, given in Table~IV, where again both $E_{0\mathrm{S}}$ as 
well as $E_{0\mathrm{T}}$, are given. In this occasion, the better compatibility of MAE and ME values signify that the error 
is systematic in nature in case of LC-BLYP. But these are always overestimated and the extent is larger than B3LYP. This may arise due to 
a greater involvement of $\gamma$ on \emph{system size}; hence possibly requires an optimally tuned $\gamma$ in theses cases. This demands 
further elaborate investigation, and may be considered in future works.  

\section{Future and Outlook}
We have examined the feasibility and practicability of a simple yet accurate time-independent DFT approach for realistic calculation 
of single excitation energies through DFT in CCG. This was applied to a host 
of organic chromophores and linear acenes having $\pi$ network. The excitation energies were estimated using two representative set of 
functionals--each from respective hybrid and RS hybrid within a pseudopotential approximation. The obtained results for all 
these species, from virial theorem are in quite good agreement with the reference results. It can be easily extended to the excitation 
energies in charge transfer complexes \citep{becke18c,feng18} as well. Its application with RS hybrid, hyper, as well as local hybrid XC 
functionals would further enhance the scope in a wide range of systems. It may also be desirable to 
examine its performance in various excited configurations other than the lowest excited state. But this may pose some computational 
challenge regarding the state-specific SCF convergence. In such occasions, some solution, as suggested in \citep{gilbert08,barca14,shea18,hait20,carter20}, 
may offer 
some valuable guidelines. 
To conclude, the present work has demonstrated the usefulness of a simple scheme in predicting optical gap in organic 
chromophores within a CCG-DFT.   
 
\section{Acknowledgement}
AG is grateful to UGC for a senior research fellowship. TG acknowledges INSPIRE program for financial support. 
AKR thankfully acknowledges funding from DST SERB, New Delhi, India (sanction order: EMR/2014/000838).

\bibliography{dftbib.bib}

\end{document}